\documentclass[aps,pra,10pt,twocolumn,amsmath,amssymb,floatfix,
longbibliography,superscriptaddress,nofootinbib]{revtex4-1}

\usepackage{amsmath,amssymb,amsfonts}
\usepackage{nicefrac}
\usepackage{graphicx}
\usepackage{hhline}
\usepackage[utf8]{inputenc}
\usepackage{subfigure}
\usepackage{xcolor}
\usepackage{hyperref}
\hypersetup{
    colorlinks=true,       
    linkcolor=blue,          
    citecolor=blue,       
    filecolor=magenta,      
    urlcolor=blue          
}

\newcommand{\eqlab}[1]{\label{eq:#1}}
\newcommand{\seclab}[1]{\label{sec:#1}}
\newcommand{\figlab}[1]{\label{fig:#1}}

\newcommand{\sref}[1]{Sec.~\ref{sec:#1}}

\newcommand{\fref}[1]{Fig.~\ref{fig:#1}}

\newcommand{\ffrefs}[2]{Figs.~\ref{fig:#1} and~\ref{fig:#2}}
\newcommand{\Fref}[1]{Figure~\ref{fig:#1}}

\newcommand{\beq}{\begin{equation}}
\newcommand{\eeq}{\end{equation}}
\newcommand{\bea}{\begin{eqnarray}}
\newcommand{\eea}{\end{eqnarray}}
\newcommand{\eq}[1]{Eq.~(\ref{eq:#1})}

\def\ip#1#2{\langle#1\vert#2\rangle}
\def\me#1#2#3{\langle#1\vert#2\vert#3\rangle}

\def\dag{^\dagger}

\def\Tr{{\rm Tr\,}}

\def\xhat{\hat{\bf x}}
\def\yhat{\hat{\bf y}}
\def\zhat{\hat{\bf z}}

\def\kk{{\bf k}}
\def\rr{{\bf r}}

\def\z2{z_{\nicefrac{1}{2}}}
\def\yz{y\vert z}
\def\yozo{y_0\vert z_0}
\def\yoz2{y_0\vert\z2}
\def\zoz2{z_0\vert\z2}

\begin{document}

\title{Gapless hinge states from adiabatic pumping of axion coupling}

\author{Thomas Olsen} \email{tolsen@fysik.dtu.dk}
\affiliation{Computational Atomic-scale Materials Design (CAMD),
  Department of Physics, Technical University of Denmark, 2800
  Kgs. Lyngby Denmark}

\author{Tom\'{a}\v{s} Rauch}
\affiliation{Friedrich-Schiller-University Jena, 07743 Jena, Germany}
\affiliation{Centro de F{\'i}sica de Materiales, Universidad del
  Pa{\'i}s Vasco (UPV/EHU), 20018 San Sebasti{\'a}n, Spain}

\author{David Vanderbilt} \affiliation{Department of Physics and
  Astronomy, Rutgers University, Piscataway, New Jersey 08854-8019,
  USA}

\author{Ivo Souza} \affiliation{Centro de F{\'i}sica de Materiales,
  Universidad del Pa{\'i}s Vasco (UPV/EHU), 20018 San Sebasti{\'a}n,
  Spain} \affiliation{Ikerbasque Foundation, 48013 Bilbao, Spain}

\begin{abstract}
We demonstrate that chiral hinge modes naturally emerge in insulating
crystals undergoing a slow cyclic evolution that changes the
Chern-Simons axion angle $\theta$ by $2\pi$. This happens when the
surface (not just the bulk) returns to its initial state at the end of
the cycle, in which case it must pass through a metallic state to
dispose of the excess quantum of surface anomalous Hall conductivity
pumped from the bulk.  If two adjacent surfaces become metallic at
different points along the cycle, there is an interval in which they
are in topologically distinct insulating states, with chiral modes
propagating along the connecting hinge.  We illustrate these ideas for
a tight-binding model consisting of coupled layers of the Haldane
model with alternating parameters.  The surface topology is determined
in a slab geometry using two different markers, surface anomalous Hall
conductivity and surface-localized charge pumping (flow of
surface-localized Wannier bands), and we find that both correctly
predict the appearance of gapless hinge modes in a rod geometry.  When
viewing the axion pump as a four-dimensional (4D) crystal with one
synthetic dimension, the hinge modes trace Fermi arcs in the Brillouin
zone of the 2D hinge connecting a pair of 3D surfaces of the 4D
crystal.
\end{abstract}
\maketitle

\section{Introduction}
\seclab{intro}

The electronic states of crystalline insulators can be characterized
by certain geometric properties of the wave functions that have
measurable consequences~\cite{vanderbilt-book18}.  For example, in
one-dimensional (1D) insulators the manifold of valence states carries
a Berry phase~\cite{zak-prl89}
\beq
\gamma=\int_0^{2\pi}\Tr\left[ A^k\right]\,dk\,,
\eqlab{gamma}
\eeq
where $k$ between $0$ and $2\pi$ is the reduced wavevector in the
Brillouin zone (BZ), and the integrand is the trace of the Berry
connection matrix $A^k_{mn}=i\ip{u_{km}}{\partial_k u_{kn}}$ over the
valence bands.  Although $\Tr[A^k]$ is not invariant under gauge
transformations among the valence states, its integral $\gamma$ is
invariant modulo $2\pi$. Physically, $\gamma$ describes the electronic
contribution to the electric polarization as
\beq
P=-e\frac{\gamma}{2\pi}\,,
\eqlab{pol}
\eeq
where $e>0$. Accordingly, the bulk polarization is itself only defined
modulo $e$ in 1D~\cite{king-smith-prb93,vanderbilt-prb93}.

The quantum of indeterminacy present in \eq{gamma} allows for the
possibility of changing $\gamma$ gradually by a multiple of $2\pi$
during a slow cyclic evolution, resulting in the transport of an
integer number $C_1$ of electrons over one lattice
constant~\cite{thouless-prb83}.  If the Hamiltonian is parametrized by
an angle $\phi$, the total change in $\gamma$ over one cycle from
$\phi=0$ to $\phi=2\pi$ is given by
\beq
\Delta\gamma=\int_0^{2\pi}d\phi\int_0^{2\pi} dk\,
\Tr \left[\Omega^{\phi k}\right]=2\pi C_1\,.
\eqlab{delta-gamma}
\eeq
$C_1$ is known as the first Chern number, and when it is nonzero the
cycle is called a Thouless pump. In \eq{delta-gamma},
$\Omega^{\phi k}_{mn}=\partial_\phi A^k_{mn}-\partial_k
A^\phi_{mn}-i[A^\phi,A^k]_{mn}$ is the covariant Berry curvature
matrix of the valence bands in $(\phi,k)$ space.  When this $(\phi,k)$
space is viewed as an effective 2D momentum space, $2\pi C_1$ becomes
a quantized Berry flux through the corresponding 2D BZ.

The Chern number $C_1$ can be defined in exactly the same way for real
2D insulators, and those for which it is nonzero are known as (first)
Chern insulators, or quantum anomalous Hall insulators. Chern
insulators were introduced by Haldane using a tight-binding
model~\cite{haldane-prl88}, and have been realized experimentally in
magnetically-doped thin films~\cite{chang-science13,chang-natmater15}.
They are characterized by a quantized anomalous Hall conductivity
(AHC) of $(e^2/h)C_1$, and by the presence of $|C_1|$ chiral edge
modes crossing the bulk gap.

In 3D, the valence bands of insulating crystals carry another global
geometric property. It is known as the Chern-Simons axion angle
$\theta$, and can be expressed as~\cite{qi-prb08,essin-prl09}
\beq
\theta=-\frac{1}{4\pi}\int_{\rm BZ}\varepsilon_{abc}\Tr
\left[ A^a\partial_b A^c-i\frac{2}{3}A^aA^bA^c\right]\,d^3k\,,
\eqlab{theta}
\eeq
where each $k_a$ runs between $0$ and $2\pi$,
$\partial_a=\partial/\partial k_a$, and $A^a_{mn}$ is the
corresponding Berry connection along lattice direction $a$.  The axion
angle describes an isotropic contribution
$\alpha_{\alpha\beta}^{\rm iso}=
(e^2/h)(\theta/2\pi)\delta_{\alpha\beta}$ to the linear
magnetoelectric coupling
$\alpha_{\alpha\beta}= \partial P_\alpha/\partial B_\beta=\partial
M_\beta/\partial{\cal E}_\alpha$.  Like the Berry phase, the axion
angle is gauge invariant only modulo $2\pi$ and can change gradually
by multiples of $2\pi$ during slow cyclic evolutions. The net change
in $\theta$ over one cycle parametrized by $\phi\in[0,2\pi]$ is given
by
\beq
\Delta\theta=
\frac{1}{16\pi}\int_{\rm BZ}d^3k\int_0^{2\pi}d\phi\,
\varepsilon_{abcd}
\Tr\left[\Omega^{ab}\Omega^{cd}\right]=2\pi C_2\,,
\eqlab{delta-theta}
\eeq
where the indices run over $k_x,k_y,k_z,\phi$.  The integer $C_2$ is
called the second Chern number, and when it is nonzero the cycle is
referred to as an axion pump~\cite{taherinejad-prl15}. Like $2\pi C_1$
given by \eq{delta-gamma}, $2\pi C_2$ can be viewed as a quantized
Berry flux through an effective BZ, which is now that of a parent 4D
insulator.

In this work, we ask what general features one can expect to see in
the boundary spectrum during an axion pumping cycle when both the bulk
{\it and} the surface return to their initial states at the end of the
cycle.  The evolution of the surface spectrum under these
circumstances was studied previously~\cite{olsen-prb17}. In the
present work we turn our attention to the hinge band structure, that
is, the spectrum of 1D modes localized at the boundaries between
contiguous surface facets.

We find that chiral hinge states appear generically in the course of
cyclic evolutions characterized by a nonzero $C_2$ invariant. In
contrast to the hinge states in intrinsic higher-order topological
insulators~\cite{Schindler2018a,geier-prb18}, their occurrence does
not rely on the presence of certain bulk crystallographic symmetries,
but only on the global Chern topology of the pumping cycle.

The paper is organized as follows. In \sref{qualitative} we provide a
qualitative discussion of the main ideas. We then illustrate them for
a concrete tight-binding model in \sref{numerics}, where we use
various tools to predict from slab calculations the occurrence of
gapless hinge modes, and to illuminate the concept of surface
topology. We conclude in \sref{summary} with a summary and outlook.

\section{Qualitative discussion}
\seclab{qualitative}

\begin{figure*}
\centering
\includegraphics[width=6.5in]{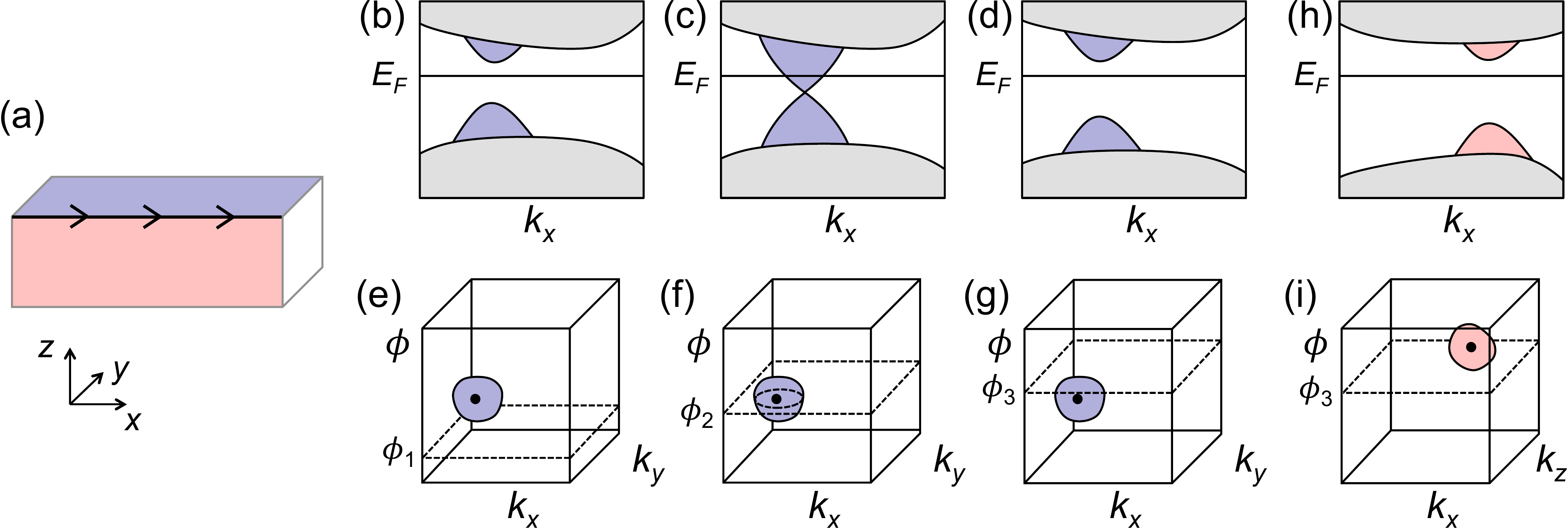}
\caption{(a) Real-space view of a sample of an insulating material
  undergoing an axion pumping cycle parametrized by $\phi$.  The
  sample is obtained by terminating a bulk crystal at two
  semi-infinite surfaces oriented normal to $+\zhat$ and $-\yhat$
  (blue and red shadings, respectively), meeting at an $x$-directed
  hinge that may harbor chiral modes (arrows).  (b-d) Hinge-projected
  band structure focusing on surface states on the $+\zhat$-oriented
  surface (blue shading), for three increasing values of $\phi$.  Gray
  shading represents projected bulk states.  (e-g) View in
  $(k_x,k_y,\phi)$ space, for the choice of Fermi level indicated in
  (b-d); blue indicates the ``Fermi surface'' in $(k_x,k_y,\phi)$
  space enclosing the electron pocket; solid dot is the Weyl point,
  corresponding to the nodal touching of surface valence and
  conduction bands at the critical parameter value $\phi_2$.  The
  dashed lines in (e-g) indicate the $\phi$ values used in (b-d),
  respectively.  (h,i) Same as (d,g), but focusing on states localized
  on the surface normal to $-\yhat$.}
\figlab{fig1}
\end{figure*}

In this section we examine the generic behavior of boundary states
during an axion pumping cycle, making no special assumptions about the
presence of symmetries or the position of the Fermi level $E_{\rm F}$
in the gap, and ask what general features one can expect to see in the
surface and hinge band structures during an axion pump evolution
cycle.

We consider for simplicity an orthorhombic structure with primitive
lattice vectors along the Cartesian axes, and work with reduced
wavevectors $(k_x,k_y,k_z)$ with each $k_j$ between 0 and $2\pi$.  To
these we can add the adiabatic Hamiltonian parameter $\phi$ whose
evolution from 0 to $2\pi$ controls the axion pump, so that we can
also think in terms of a 4D insulator with a second Chern number $C_2$
in $(k_x,k_y,k_z,\phi)$ momentum space. In the present discussion we
shall assume $C_2=1$, so that one quantum of axion coupling is pumped
during the adiabatic cycle.

\subsection{Surface states}
\seclab{surface-states}

Consider the system shown in \fref{fig1}(a) -- a crystal terminated at
two semi-infinite surfaces normal to $+\zhat$ and $-\yhat$, meeting at
an $x$-directed hinge.  In preparation for the discussion of
hinge-localized states in the next subsection, here we consider the
band structures of the two surfaces, but projected as though seen from
the hinge.

We focus first on the top surface (unit normal $+\zhat$).  The
evolution of its hinge-projected band structure is sketched in panels
(b-d) for three increasing values of $\phi$ in the region where the
metalization occurs.  These are labeled as $\phi_1$, $\phi_2$, and
$\phi_3$, corresponding to panels (e-g) respectively, where the locus
of electron-occupied surface states is indicated in the 3D
$(k_x,k_y,\phi)$ space.  As a reminder, the surface is required to
become metallic over some range of $\phi$. This follows because we
assume that the surface Hamiltonian (as well as the bulk one) returns
to itself at the end of the $\phi$ loop, so that the quantum of AHC
that is pumped to this surface has to be removed by a metallic
interval~\cite{olsen-prb17}.  Typically this happens as shown in
panels (b-d). That is, surface states penetrate into the gap with
increasing $\phi$, leading to the formation of a nodal touching in
$(k_x,k_y)$ space at the critical parameter value $\phi_2$, after
which the gap reopens to restore an insulating surface. When viewed in
$(k_x,k_y,\phi)$ space, that nodal touching becomes a Weyl point,
indicated by the solid dot in panels (e-g).

If one would follow the evolution of the surface AHC by computing the
contributions only up to the nodal point, one would observe a sudden
jump by $e^2/h$ when passing through $\phi_2$.  This jump is precisely
by the amount needed to return the surface AHC to its initial value at
the end of the cycle.  With the indicated Fermi-level position,
however, the change occurs continuously.  An electron pocket first
appears when the conduction band minimum drops below $E_{\rm F}$,
somewhere between panels (b) and (c); it grows, then shrinks and
disappears somewhere between (c) and (d). This behavior is visualized
in 3D $(k_x,k_y,\phi)$ space in panels (e-g), with the dashed
rectangles showing the $\phi$ values corresponding to panels (b-d),
respectively.

At any value of $\phi$ for which the surface electron pocket exists,
such as that shown by the horizontal cut in \fref{fig1}(f), the
contribution of that pocket to the surface AHC is proportional to the
Berry phase computed around its boundary.  This phase evolves by
$2\pi C_1^{\text{FS}}$ from the creation to the destruction of the
pocket, where $C_1^{\text{FS}}$ is the first Chern number (one, in our
case) on the spheroidal Fermi surface shown in panels (e-g) of
\fref{fig1}.

We can regard any one of those three panels as showing the
Fermi-surface structure of the 3D ($k_x,k_y,\phi$) system
corresponding to the $z$-terminated 3D surface of a 4D
($k_x,k_y,k_z,\phi$) second-Chern insulator.  We see a Fermi pocket
with nonzero first Chern index surrounding a Weyl node, shown as a
dark central point in each panel.  This looks very much like a picture
of a Weyl semimetal~\cite{armitage-rmp18}, but with one crucial
difference. In a true 3D system, the Nielsen-Ninomiya
theorem~\cite{nielsen-plb83} requires that the chiralities of the Weyl
nodes must sum to zero over the 3D BZ. The violation we see here is an
example of an anomaly; since we are at the surface of a topological 4D
insulator, the reasoning used to prove the Nielsen-Ninomiya theorem no
longer applies. In fact, the sum of chiralities is necessarily equal
to the 4D bulk second Chern number, that is, $C_1^{\text{FS}}=C_2$.
Hence, every 3D surface facet of a 4D second-Chern insulator must show
the same excess of chirality~\cite{qi-prb08,olsen-prb17}.  In
particular, every 3D surface must be metallic, in analogy to the 1D
surfaces of a 2D first-Chern insulator.

Panels~(h) and (i) of \fref{fig1} show similar plots at the same
$\phi=\phi_3$ value as in panels (d) and (g), but now for the surface
with unit normal $-\yhat$.  Here we assume that the metallic interval
has not yet begun, so the electron pocket corresponding to this
surface lies above the $\phi=\phi_3$ plane.  Incidentally, if the
Fermi-level position had been chosen lower in the gap, the metallic
interval of $\phi$ could correspond to the temporary creation of a
hole pocket instead, on either or both of the surface facets; entirely
parallel arguments apply in these cases.

\begin{figure}[b]
\centering
\includegraphics[width=\columnwidth]{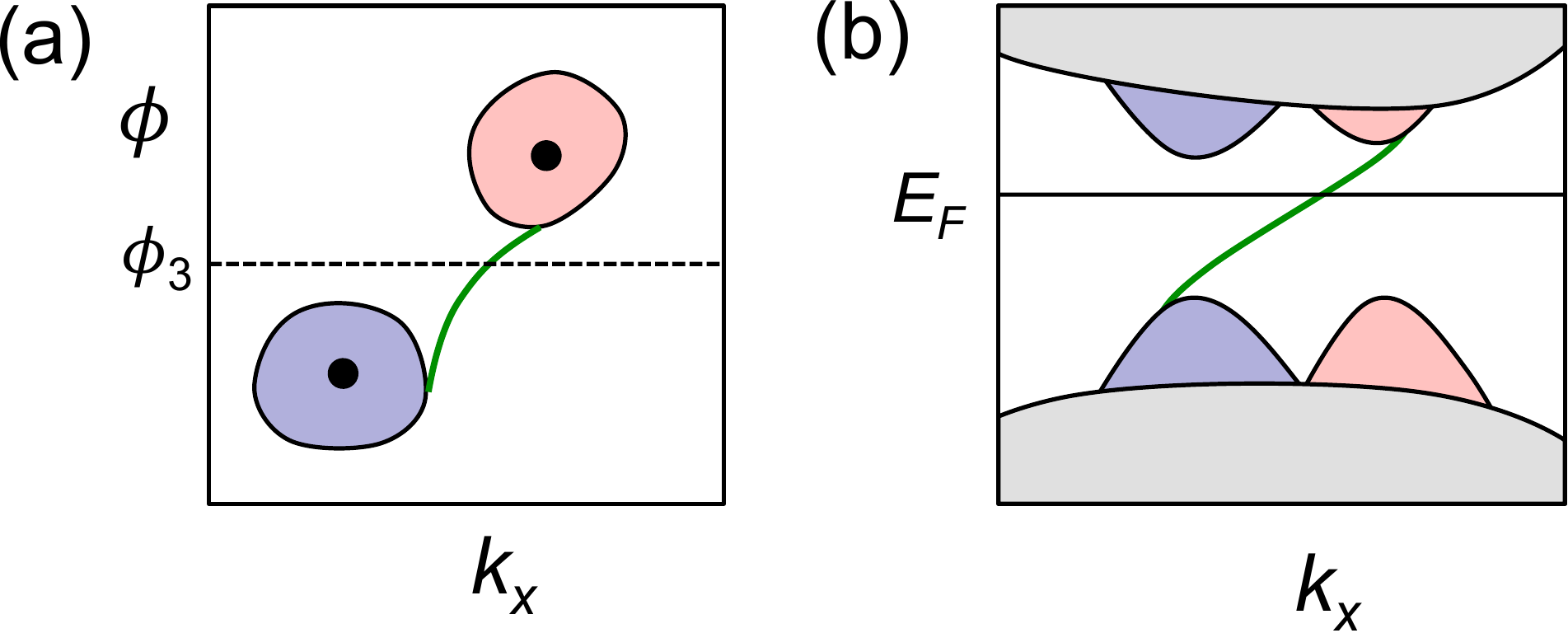}
\caption{(a) Hinge-projected Fermi surface of the system depicted in
  \fref{fig1}(a), plotted in $(k_x,\phi)$ space. Blue and red electron
  pockets correspond to surface states on the $+\zhat$- and
  $-\yhat$-oriented surfaces, as shown in \fref{fig1}(g) and (i),
  respectively. Fermi-arc state is shown in green.  (b)
  Hinge-projected band structure plot vs.~$k_x$ (at $\phi=\phi_3$,
  indicated by the dashed line in (a)). The gray, blue, and red
  regions are the projected bulk, $+\zhat$-surface, and
  $-\yhat$-surface states shown in Figs.~\ref{fig:fig1}(d,f), and the
  green line is the chiral hinge state.}
\figlab{fig2}
\end{figure}

\subsection{Hinge states}
\seclab{hinge-states}

Now consider the $x$-directed hinge adjoining the $y$- and
$z$-oriented surface facets discussed above (we refer to it as a
``$\yz$ hinge''). \Fref{fig2}(a) shows the locus of points on the
$(k_x,\phi)$ plane where there are states at $E_{\rm F}$, that is, the
hinge-projected Fermi surface plotted as a function of $\phi$.  The
blue region is the projection onto the $(k_x,\phi)$ plane of the Fermi
surface in $(k_x,k_y,\phi)$ space of the $+\zhat$-oriented surface
[\fref{fig1}(g)]. Likewise, the red region is the projection of the
$(k_x,k_z,\phi)$ Fermi surface of the $-\yhat$-oriented surface
[\fref{fig1}(i)] adjoining the $+\zhat$ surface at the hinge.  Each
electron pocket encloses a Weyl point, and these bring opposite chiral
charges due to the fact that positive circulations on the $+\zhat$-
and $-\yhat$-oriented surfaces correspond to positive and negative
transport respectively at the hinge.

The Nielsen-Ninomiya theorem is now satisfied, since the total
chirality of all Weyl points projecting onto the $(k_x,\phi)$ plane
necessarily vanishes.  However, the separation of chiral charges
between the two pockets requires the presence of a Fermi arc
connecting them, as shown by the green line in \fref{fig2}(a), just as
for the case of a Fermi arc at the surface of a Weyl
semimetal~\cite{armitage-rmp18}.  Indeed, the count of Fermi arcs and
geometry of attachment must follow the same rules outlined by Haldane
\cite{haldane-arxiv14}.  The full hinge-projected band structure at
$\phi=\phi_3$ is illustrated in \fref{fig2}(b), showing the
hinge-localized state that crosses from the valence to the conduction
manifold and gives rise to the Fermi arc.

Of course, at other points along the adiabatic pumping cycle,
corresponding to different values of $\phi$, the chiral hinge mode may
be absent, as above the red region or below the blue region in
\fref{fig2}(a).  At some other $\phi$ values, the presence of hinge
states will be obscured by degeneracy with the continuum of surface
states.  In some cases these regions of metallic surface behavior
could be much more extensive than as sketched above, hiding the hinge
electronic structure almost completely. However, we can be sure of the
existence of at least one Weyl-node-surrounding hole or electron
pocket arising from each adjoined surface, and a required Fermi arc
state connecting them somewhere in the $(k_x,\phi)$ space (unless the
two pockets overlap when projected into this space).  In this sense,
the presence of Fermi arcs and chiral hinge channels at some stage of
a second-Chern pumping cycle is generic.

In any case, these observations all serve to illustrate the very close
analogy between the physics on the 2D ``hinge'' of a 4D second-Chern
topological insulator, and at the 2D surface of a 3D Weyl semimetal.

\section{Numerical study of a toy model}
\seclab{numerics}

\subsection{The alternating Haldane model}
\seclab{model}

To illustrate the physics described above, we study the tight-binding
model for an axion pump introduced in Ref.~\cite{olsen-prb17},
consisting of alternating layers of the Haldane
model~\cite{haldane-prl88}. The on-site energies are modulated by an
angle $\phi$ in such a way that for $-\pi/2<\phi<\pi/2$ the first
Chern numbers vanish on all layers, while for $\pi/2<\phi<3\pi/2$ they
alternate between $+1$ and $-1$, for isolated layers.  To prevent the
layers from becoming metallic at $\phi=\pi/2$ and $3\pi/2$,
$\phi$-dependent interlayer couplings are introduced.  As $\phi$ goes
from $0$ to $2\pi$, the system is carried along a gapped circuit that
encloses a gapless point in parameter space, and the axion angle
$\theta$ increases gradually from $0$ to $2\pi$~\cite{olsen-prb17}.

When viewed along the stacking direction $z$, the model consists of
coupled chains that project onto the honeycomb sites on each layer,
with alternating on-site energies and hoppings along $z$.  In the
limit of vanishing interchain coupling, the chain Hamiltonian is
identical to the Rice-Mele model of alternating site energies and
hopping strengths~\cite{rice-prl82}, which realizes a Thouless
pump~\cite{vanderbilt-prb93}. Chains passing through the $A$ and $B$
sites have equal and opposite first Chern numbers in $(\phi,k_z)$
space, so that no net charge is transported along $z$ over one
cycle. Depending on the choice of parameters, the magnitude of those
Chern numbers is either zero or one.

The 2D unit cell of each layer is spanned by the lattice vectors
${\bf a}_1=a\xhat$ and ${\bf a}_2=a\xhat/2+\sqrt{3}a\yhat/2$, with
orbitals sitting on the honeycomb sites
${\bf t}_{\rm A}={\bf a}_1/3+{\bf a}_2/3$ and
${\bf t}_{\rm B}=2{\bf a}_1/3+2{\bf a}_2/3$. The Hamiltonian for an
isolated layer indexed by $p$ is
\beq
\begin{aligned}
H_p&=(-1)^p\Delta\sum_{i}\gamma_i c_{pi}\dag c_{pi}
+t\sum_{\langle ij\rangle}c_{pi}\dag c_{pj}\\
+&(-1)^p
\sum_{\langle\kern-2.0pt\langle ij\rangle\kern-2.0pt\rangle}i\nu_{ij}
c_{pi}\dag c_{pj}\,,
\end{aligned}
\eqlab{H-monolayer}
\eeq
where $i$ and $j$ label the sites, with $\gamma_i=\pm 1$ if site $i$
belongs to the A or B sublattice. $\langle ij\rangle$ and
$\langle\kern-2.0pt\langle ij\rangle\kern-2.0pt\rangle$ denote pairs
of first and second nearest-neighbor sites, with each pair appearing
twice. The first and second terms contain the on-site energies and
nearest-neighbor hoppings respectively, and the third describes a
pattern of staggered magnetic fluxes generated by complex
second-neighbor hoppings of unit magnitude. Therein, $\nu_{ij}=+1$
($-1$) if the hopping direction from $j$ to $i$ is right-handed
(left-handed) around the center of a plaquette.  The $(-1)^p$ factor
in the first term reverses the energies of sites on the same
sublattice in adjacent layers, while the same factor in the third term
reverses the pattern of magnetic fluxes, and with it the first Chern
numbers on consecutive layers.  The hopping magnitude in the third
term has been set to unity as a reference, and each 2D layer undergoes
a Chern transition between topological and trivial phases at
$\Delta=\pm3\sqrt{3}$.  At $\Delta=3\sqrt{3}$ the gap-closing
transition occurs at the high-symmetry point
$\overline{K}=(4\pi/3a)\xhat$ in the 2D BZ, and at $\Delta=-3\sqrt{3}$
it occurs at $\overline{K^{\prime}}=-(4\pi/3a)\xhat$.

The full 3D model has ${\bf a}_3=c\zhat$ as the third lattice vector,
and two layers per unit cell.  The layers $p=0,1$ are located at
$z=-c/4,c/4$, and the Hamiltonian reads
\beq
 H_{\rm bulk}=\sum_p
\bigg[
H_p + \Big[1+(-)^p\, t'\Big]
\sum_i\gamma_i\left(c_{pi}\dag c_{p+1,i}+\text{H.c.}\right)
\bigg]\,,
\eqlab{bulk}
\eeq
where the second term describes the interlayer (intrachain) coupling
and ``H.c'' stands for ``Hermitian conjugate.''  We choose $t=-4.0$,
and parameterize $\Delta$ and $t^\prime$ according to
\begin{subequations}
\begin{align}
\Delta&=3\sqrt{3}+2\cos\phi\,,
\eqlab{Delta}\\
t^\prime&=0.4\sin\phi\,.
\eqlab{t-int-bulk}
\end{align}
\eqlab{circuit}
\end{subequations}
The presence of a nonzero $t^\prime$ introduces an alternation of
interlayer hopping strengths that keeps the system gapped as $\phi$
passes through $\pi/2$ and $3\pi/2$ where the topological transitions
occur in the isolated layers.  The bulk spectrum is therefore gapped
everywhere along the adiabatic cycle parameterized by $\phi$,
encircling a gapless point at $(\Delta,t^\prime)=(3\sqrt{3},0)$.

The model has neither time reversal (TR) nor inversion symmetry at
generic $\phi$.  However, the Hamiltonian is invariant under an
antiunitary operator
\beq
\Lambda = K i\tau_y\sigma_x = K \tau_z I
\eqlab{trs}
\eeq
where $K$ is complex conjugation and $\tau_j$ and $\sigma_j$ are the
$j$'th Pauli matrices acting in the layer and sublattice spaces
respectively.\footnote{We thank N.~Varnava for pointing out this
  symmetry of the model.}
Inversion about a hexagonally centered point midway between the layers
is represented by $I=\tau_x\sigma_x$, so the second equality expresses
$\Lambda$ as $I$ followed first by a sign reversal of all amplitudes
on odd layers, then by scalar TR.  Because $\Lambda$ is antiunitary
and squares to $-1$, it acts the way inversion times TR does in a
spinor system, forcing the four energy bands to come in two
Kramers-degenerate pairs.  At the points $\phi=0$ and $\pi$ where the
alternation $t^\prime$ of the interlayer hoppings vanishes, the model
acquires two additional symmetries: mirror symmetry $M_z$
($z\rightarrow -z$) about the layers, and time-reversal $K$ combined
with a twofold rotation $C_2^y$ about an axis lying on the atomic
layers and pointing along the armchair edges.

\begin{figure}[t]
\centering
\includegraphics[width=0.7\columnwidth]{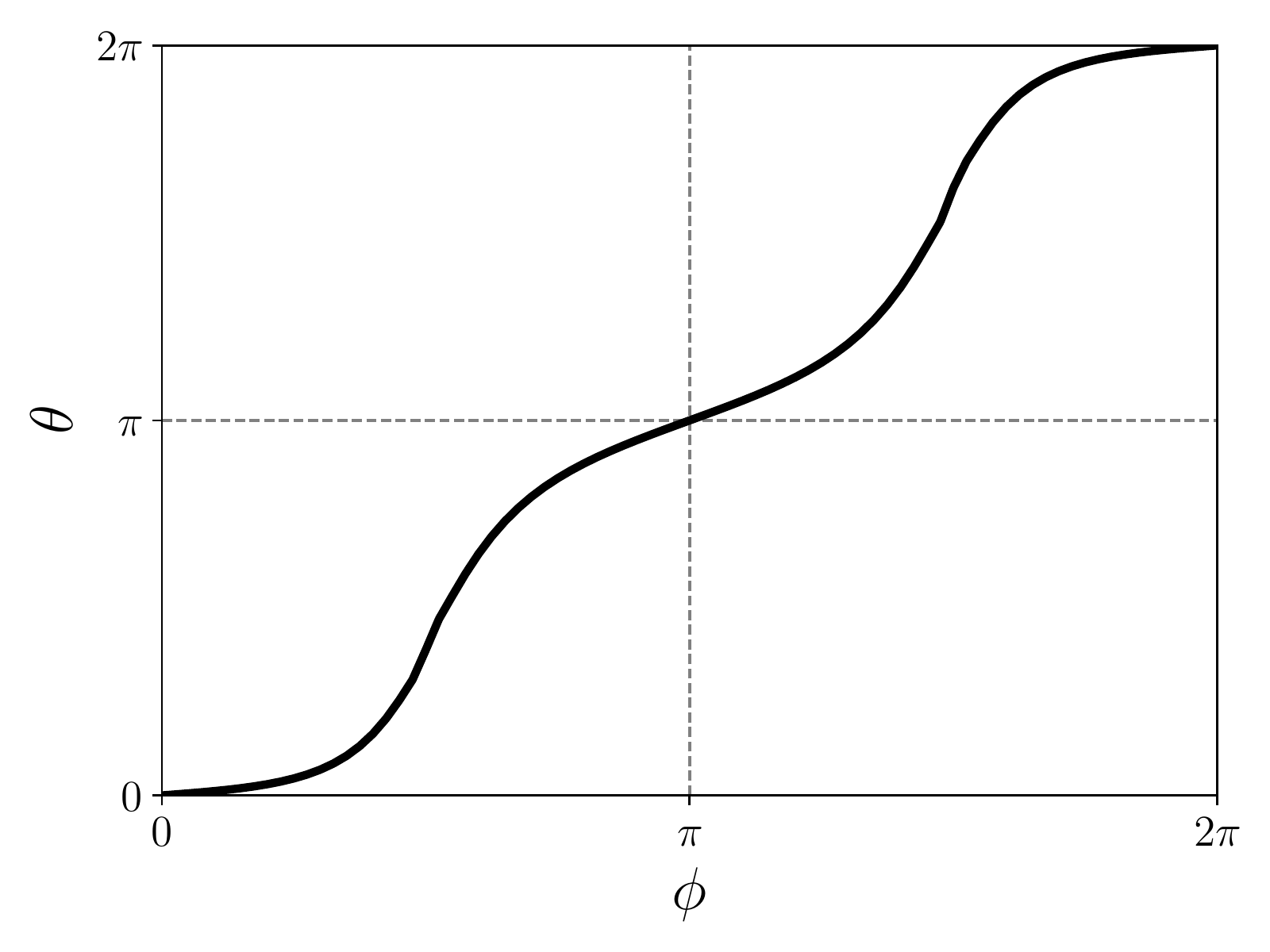}
\caption{Pumping of the axion angle $\theta$ by $2\pi$ in the
  alternating Haldane model.}
\figlab{fig3}
\end{figure}

The evolution with $\phi$ of the axion angle is shown in
\fref{fig3}. $\theta$ increases gradually from $0$ to $2\pi$ over one
cycle, corresponding to $C_2=1$ in \eq{delta-theta}.  Both $M_z$ and
$KC_2^y$ take $\theta$ into $-\theta$, constraining $\theta$ to be
$0\text{ mod }\pi$ at $\phi=0$ and $\pi$, consistent with the figure.
At $\phi=0$ the system is a topologically trivial insulator with
$\theta=0$. Instead, at $\phi=\pi$ it is a topological crystalline
insulator with $\theta=\pi$ (a ``generalized axion insulator'' in the
sense of Ref.~\cite{varnava-prb20}), harboring metallic states on
surfaces that preserve either $M_z$ or $KC_2^y$ symmetry, or
both~\cite{olsen-prb17,varnava-prb20}.

\subsection{Surface topological transitions and surface anomalous Hall
  conductivity}
\seclab{surf-AHC}

\begin{figure}
\centering
\includegraphics[width=0.99\columnwidth]{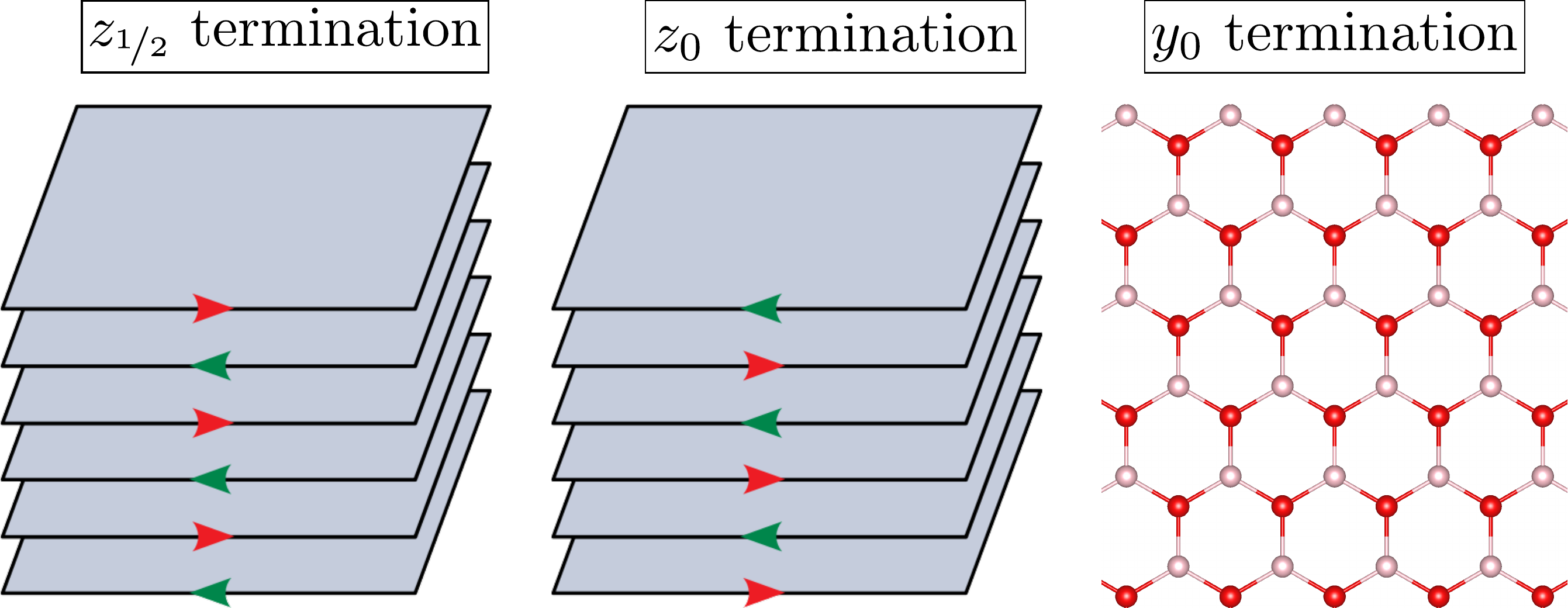}
\caption{Left and middle: inequivalent $z$-terminated slabs of the
  alternating Haldane model.  Arrows indicate edge-mode chiralities on
  the uncoupled layers for $\pi/2<\phi<3\pi/2$.  Right: one layer of a
  $y_0$-terminated slab, with zigzag edges (top view).}
\figlab{fig4}
\end{figure}

We study three types of slabs, shown schematically in \fref{fig4}. The
ones in the left and middle panels are terminated along $z$, and the
one on the right is terminated along $y$. In the left panel the
stacking unit is a cell with boundaries at $z=\pm 1/2$ (in units of
$c$), enclosing layers located at $z=\pm 1/4$.  With this ``$\z2$
termination,'' the Chern numbers of the top and bottom layers (when
isolated) are $C_1=\pm 1$ respectively in the interval
$\pi/2<\phi<3\pi/2$.  In the middle panel the stacking unit is a cell
with boundaries at $z=0$ and $1$ and enclosing layers at $z=1/4$ and
$3/4$.  With this ``$z_0$ termination,'' the top and bottom layers
have Chern numbers $C_1=\mp 1$ respectively in the same interval.
Finally, in the right panel we have a slab with a ``$y_0$
termination'' consisting of zigzag edges on every layer.

For each type of slab we perform the cyclic evolution described by
\eq{circuit}, with the surfaces returning to their initial states
along with the bulk.  The surfaces must then pass through metallic
states to dispose of the quantum of surface AHC pumped from the bulk.
To visualize the gap closure, we plot in \fref{fig5}(a) the minimum
energy gap as a function of $\phi$. There is one gap closure per
cycle, as in \fref{fig1}(c), taking place at isolated critical values
$\phi_c$ that are different for the three slabs.  An examination of
the slab band structures~\cite{olsen-prb17} reveals that at $\phi_c$
the valence and conduction surface bands touch at a nodal point, which
occurs at precisely $E_{\rm F}$ because we consider the slabs at half
filling.  If we were to shift $E_{\rm F}$ away from the nodal point as
in \fref{fig1}(c), each surface would remain metallic over a finite
$\phi$ interval containing~$\phi_c$, as illustrated in
\fref{fig1}(e-g).

\begin{figure}[tb]
\centering
\includegraphics[width=0.85\columnwidth]{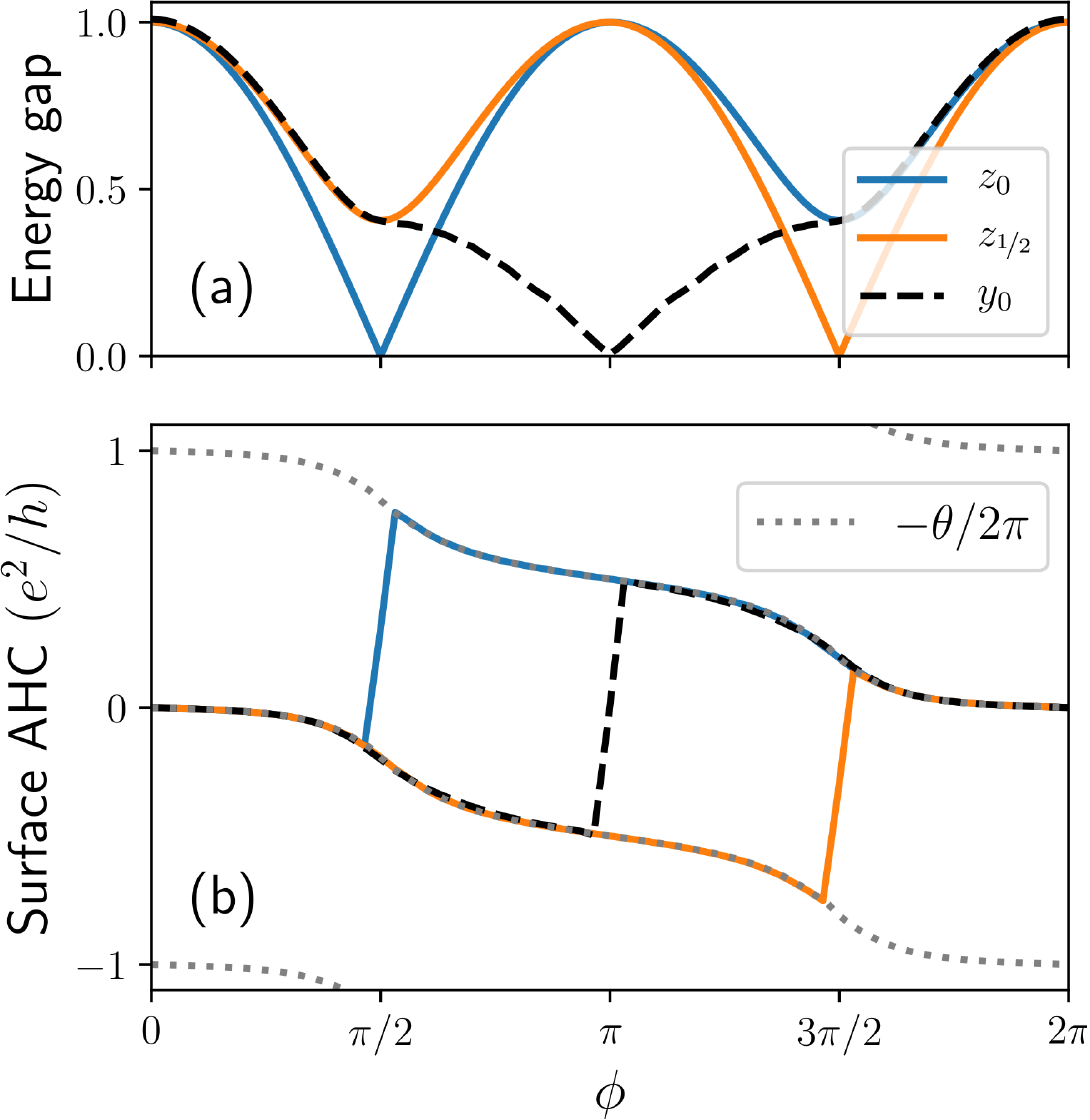}
\caption{(a) Evolution of the minimum energy gap versus $\phi$ in
  slabs of the alternating Haldane model, for the three types of
  surface terminations pictured in \fref{fig4}. The gap closures occur
  at the surfaces. (b) Evolution of the surface AHC of each
  slab. Dotted lines are different branches of the bulk axion
  coupling, plotted as $-\theta/2\pi$ according to \eq{surf-ahc}. The
  finite slopes of the discrete jumps at $\phi=\pi/2$, $\pi$, and
  $3\pi/2$ are artifacts of the finite step size used for $\phi$ in
  the calculation.}
\figlab{fig5}
\end{figure}

The gap closure is pinned to $\phi_c=\pi$ on $y_0$ surfaces, because
at $\phi=\pi$ the system becomes a generalized axion insulator
protected by $M_z$ and $KC_2^y$ symmetries, both of which are
preserved at those surfaces.\footnote{The symmetry $M_z$ (but not
  $KC_2^y$) is also preserved at $x$-oriented surfaces terminated at
  armchair edges, and this suffices to pin the gap closure to
  $\phi_c=\pi$ on those surfaces as well.}
The gap closing occurs exactly at $E=0$ as an artifact of a
particle-hole symmetry in the model.  As for the $z$-oriented
surfaces, neither $M_z$ nor $KC_2^y$ symmetry is preserved there, so
these surfaces are not required to be metallic at
$\phi=\pi$. Nevertheless they must still become metallic somewhere
along the cycle, and the closing of the gap occurs at $\phi_c=\pi/2$
on the $z_0$ surface and at $\phi_c=3\pi/2$ on the $\z2$ surface,
again exactly at $E=0$. (The closing occurs at point $\overline{K}$ in
the 2D BZ~\cite{olsen-prb17}, where A and B chains become decoupled,
with $H_{\rm A}=-H_{\rm B}$. At generic $\phi$ each takes the form of
a Rice-Mele chain~\cite{rice-prl82}, and the surface gap closure
occurs when the surface-state energies of the two chains cross through
each other and through zero, which occurs at $\cos\phi=0$ where the
effective site energy alternation vanishes.)

These gap-closing events at the surfaces are topological phase
transitions, and to elucidate the notion of surface topology we now
examine the AHC carried by the surfaces along the pumping cycle.  For
$E_{\rm F}=0$ and in the limit of a thick slab, we expect the surface
AHC to jump by $e^2/h$ at $\phi_c$, as described by the relation
\beq
\sigma_{\rm AHC}^{\rm surf}=
\left(n-\theta/2\pi\right)\frac{e^2}{h}
\eqlab{surf-ahc}
\eeq
between the AHC of a gapped surface and the bulk axion
coupling~\cite{essin-prl09,rauch-prb18}. Once a specific branch has
been chosen for $\theta$, a unique integer $n$ can be assigned to each
surface, and for $n$ to change the surface gap must close and reopen.
The difference in AHC between two insulating surface terminations of
the same bulk is $(e^2/h)\Delta n$, where $\Delta n$ is the difference
between the $n$ values on the two surfaces. In the $\phi$ intervals
where $\Delta n$ is nonzero the two surfaces are in topologically
distinct states, and if they meet there will be $|\Delta n|$ chiral
modes propagating along the adjoining hinge~\cite{sitte-prl12}.

We have calculated the surface AHC according to
Refs.~\cite{rauch-prb18,varnava-prb18} for slabs of different
thicknesses (7, 13, and 19 cells across $y$, and 7, 9, and 11 cells
across $z$).  The extrapolated results are plotted in \fref{fig5}(b),
confirming that \eq{surf-ahc} is satisfied throughout the cycle. The
AHC of each surface tracks one branch of $-\theta/2\pi$ for
$0\leq\phi<\phi_c$, switches to another branch at $\phi_c$, and
returns to its initial value at the end of the cycle.  We see that the
$y_0$ surface is topologically distinct from the $z_0$ surface for
$\phi\in(-\pi/2,\pi/2)$ and from the $\z2$ surface for
$\phi\in(\pi,3\pi/2)$, with $\Delta n=-1$ and $\Delta n=+1$
respectively.  Gapless modes are therefore expected to appear on the
$\yozo$ hinges in the former interval and on the $\yoz2$ hinges in the
latter, with opposite chiralities in the two cases. This is
illustrated by the outer and middle racetracks in the phase diagram of
\fref{fig6}.

Finally, the topological difference $\Delta n=+1$ between $z_0$ and
$\z2$ surfaces for $\phi\in(\pi/2,3\pi/2)$ can be understood as
follows.  To switch from one termination to the other one either
removes the outer surface layer, or adds an extra layer. Doing so
changes the surface AHC by $\pm e^2/h$ in the range $(\pi/2,3\pi/2)$
where the individual layers have Chern numbers $\pm 1$, and leaves the
surface AHC unchanged in the range $(-\pi/2,\pi/2)$ where the layer
Chern numbers vanish.  A similar behavior was observed in
Ref.~\cite{varnava-prb18} for a model of an axion insulator, where the
half-quantized surface AHC changed sign when a surface layer was added
or removed.  As a consequence, every single-layer-high surface step
carries a chiral edge channel in such cases~\cite{mong-prb10}, as
shown for our model by the inner racetrack in \fref{fig6}.

\subsection{Gapless hinge modes and hinge Fermi arcs}
\seclab{hinge-rod}

\begin{figure}[tb]
\centering
\includegraphics[width=0.55\columnwidth]{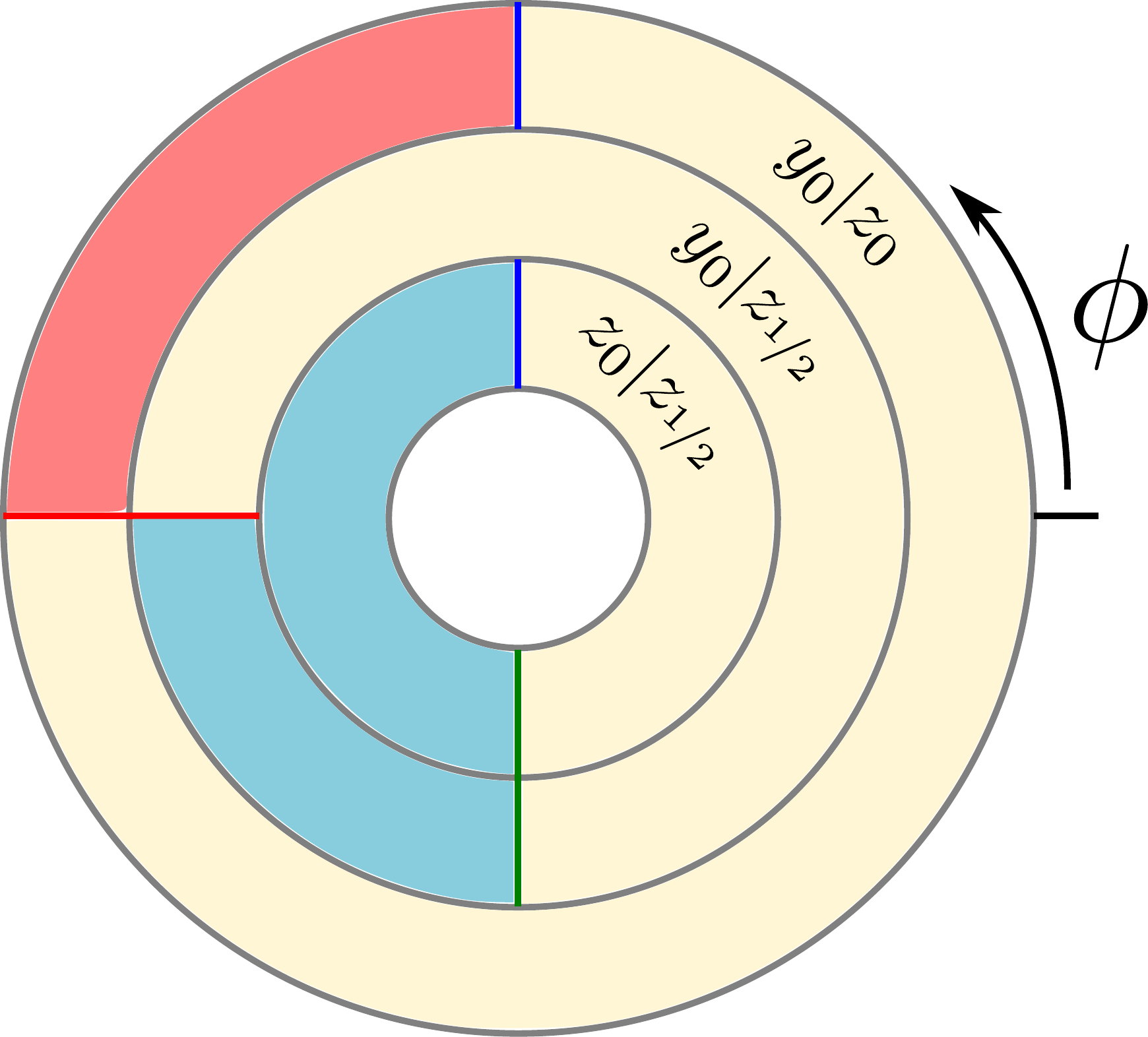}
\caption{Topological phase diagram for 1D channels in the alternating
  Haldane model at half filling. The outer and middle racetracks are
  for the two types of $\yz$ hinges, and the inner one is for
  single-layer-high steps on $z$-oriented surfaces. In the yellow
  regions there are no protected 1D modes because the surface-AHC
  difference in \fref{fig5} is $\Delta n=0$, while in the blue
  ($\Delta n=+1$) and red ($\Delta n=-1$) regions there is one
  protected mode per hinge or step. Red, blue, and green lines mark
  the gap-closing points $\phi_c$ on the $y_0$, $z_0$, and $\z2$
  surfaces, respectively, that separate the different phases.}
\figlab{fig6}
\end{figure}

\begin{figure} 
\centering
\includegraphics[width=7.0 cm]{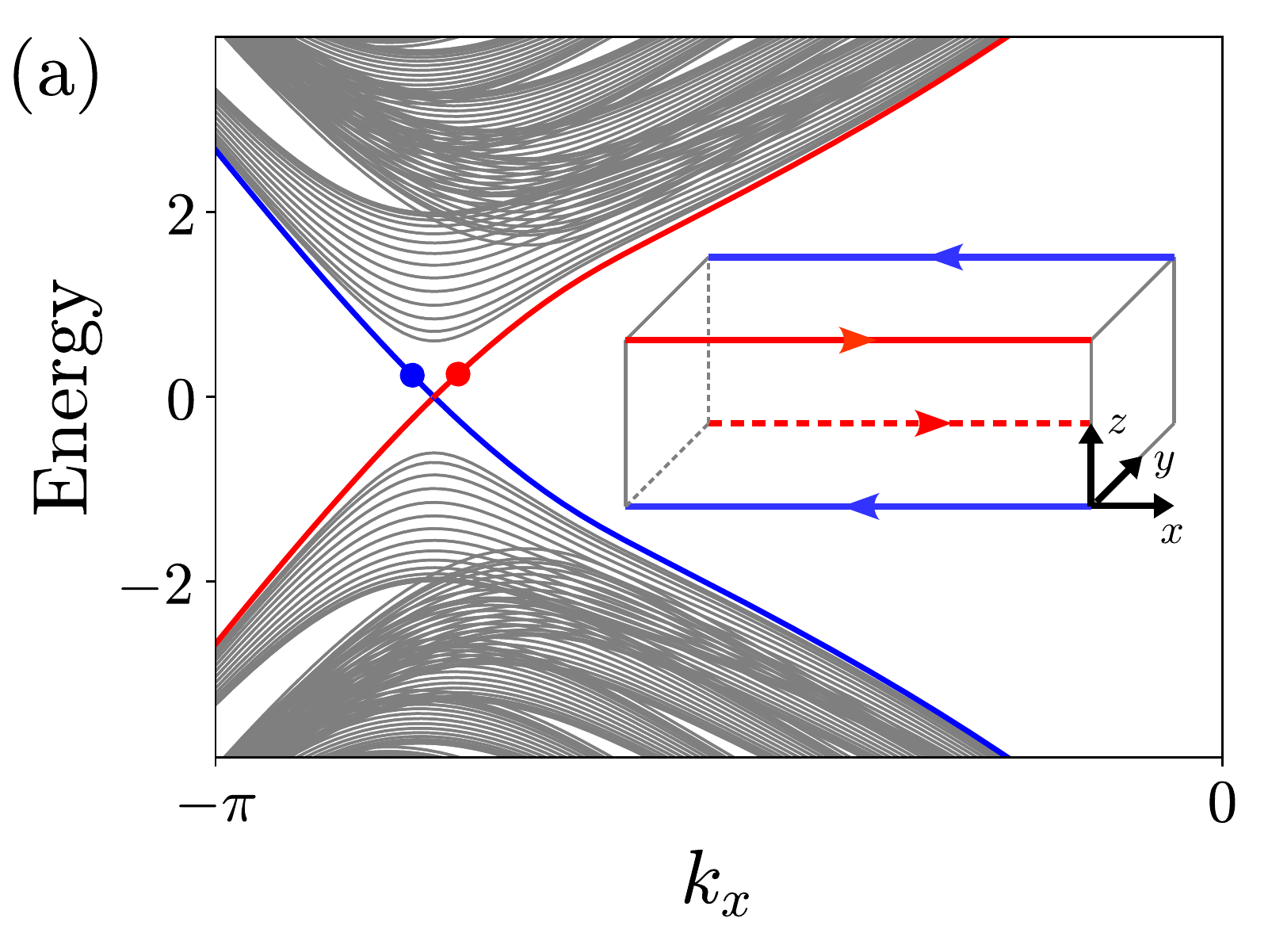}\\
\includegraphics[width=7.0 cm]{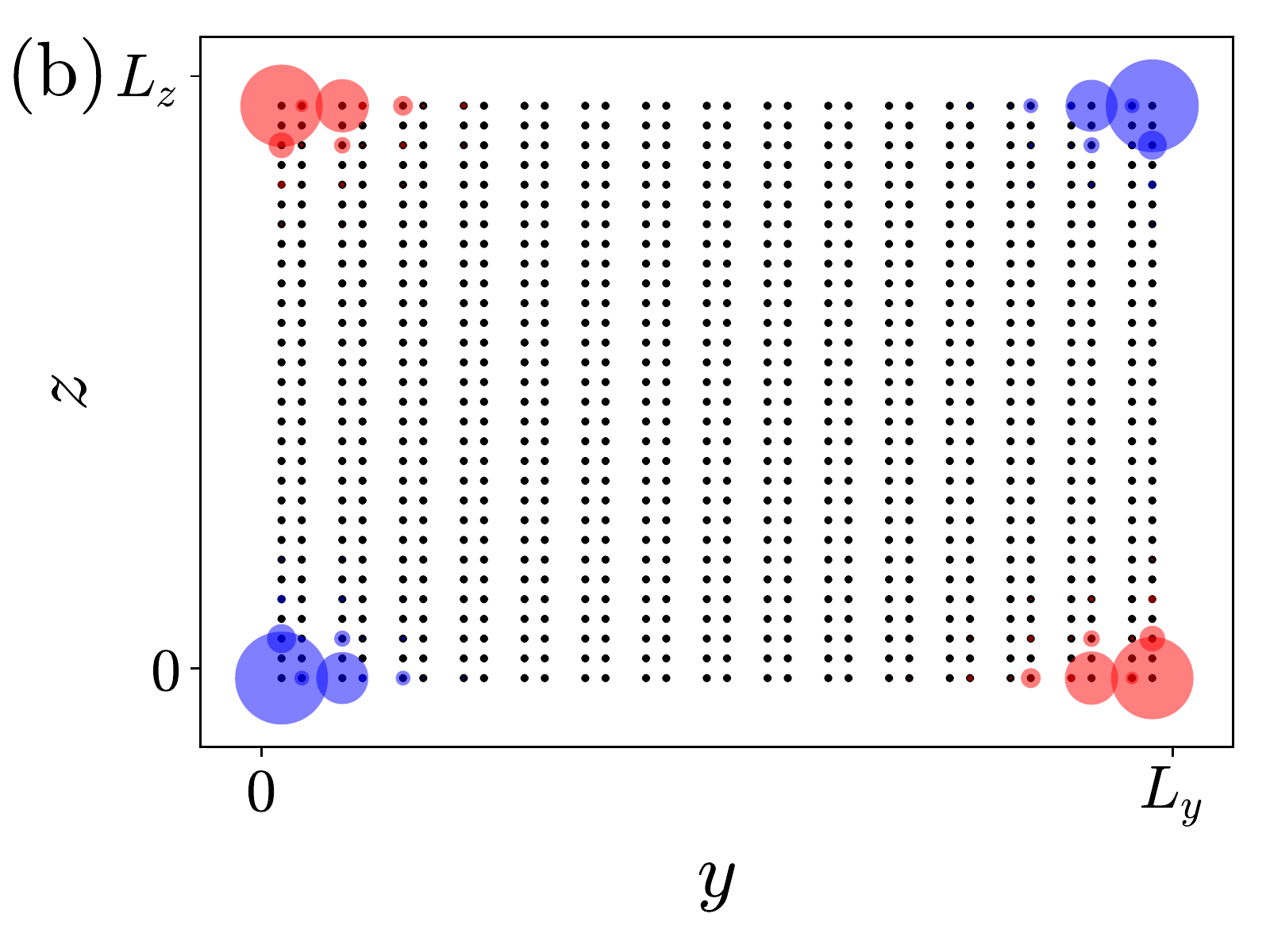}
\caption{(a) Energy bands of the alternating Haldane model, calculated
  at $\phi=5\pi/4$ for a rod extended along $x$ and with $y_0$ and
  $\z2$ terminations along $y$ and $z$.  All bands are doubly
  degenerate, and those in red and in blue are hinge-localized chiral
  modes crossing the bulk and surface gaps, depicted schematically in
  the inset.  (b) Site-resolved weights of the four hinge-localized
  states at an energy slightly above the crossing point in the middle
  of the gap, as indicated by the blue and red dots in (a).}
\figlab{fig7}
\end{figure}

To verify the presence of protected hinge modes in the predicted
$\phi$ intervals, we have studied rod-shaped samples extended along
$x$, and 15-cells thick along both $y$ and $z$.  \Fref{fig7}(a) shows
the energy bands of a $y_0$- and $\z2$-terminated rod at $\phi=5\pi/4$
(the middle of the $\phi$ interval where gapless hinge modes are
expected to occur).  All bands are doubly degenerate, since the
Kramers-enforcing operator $\Lambda$ of \eq{trs} remains a symmetry of
the rod as a whole, and the bands drawn in red and in blue are the
predicted hinge modes crossing the bulk gap.  The weights of their
wave functions on each site are displayed in \fref{fig7}(b) at an
energy near $E=0$ (the middle of the gap); modes localized on adjacent
hinges disperse in opposite directions, forming the pattern shown in
the inset of panel~(a).

The spectrum looks qualitatively the same for any value of $\phi$
between $\pi$ and $3\pi/2$; when passing through $\pi$ or $3\pi/2$,
the surface gap closes and reopens on one of the surfaces, allowing a
change of surface topology such that the band crossing on the hinge no
longer occurs.  Outside that interval, the highest-occupied and
lowest-unoccupied states become delocalized over the entire rod.  When
the surface termination is changed from $\z2$ to~$z_0$ the interval
hosting gapless modes changes from $(\pi,3\pi/2)$ to $(\pi/2,\pi)$ and
the chiralities get reversed, as predicted.

\begin{figure}[tb]
\centering
\includegraphics[width=7.6 cm]{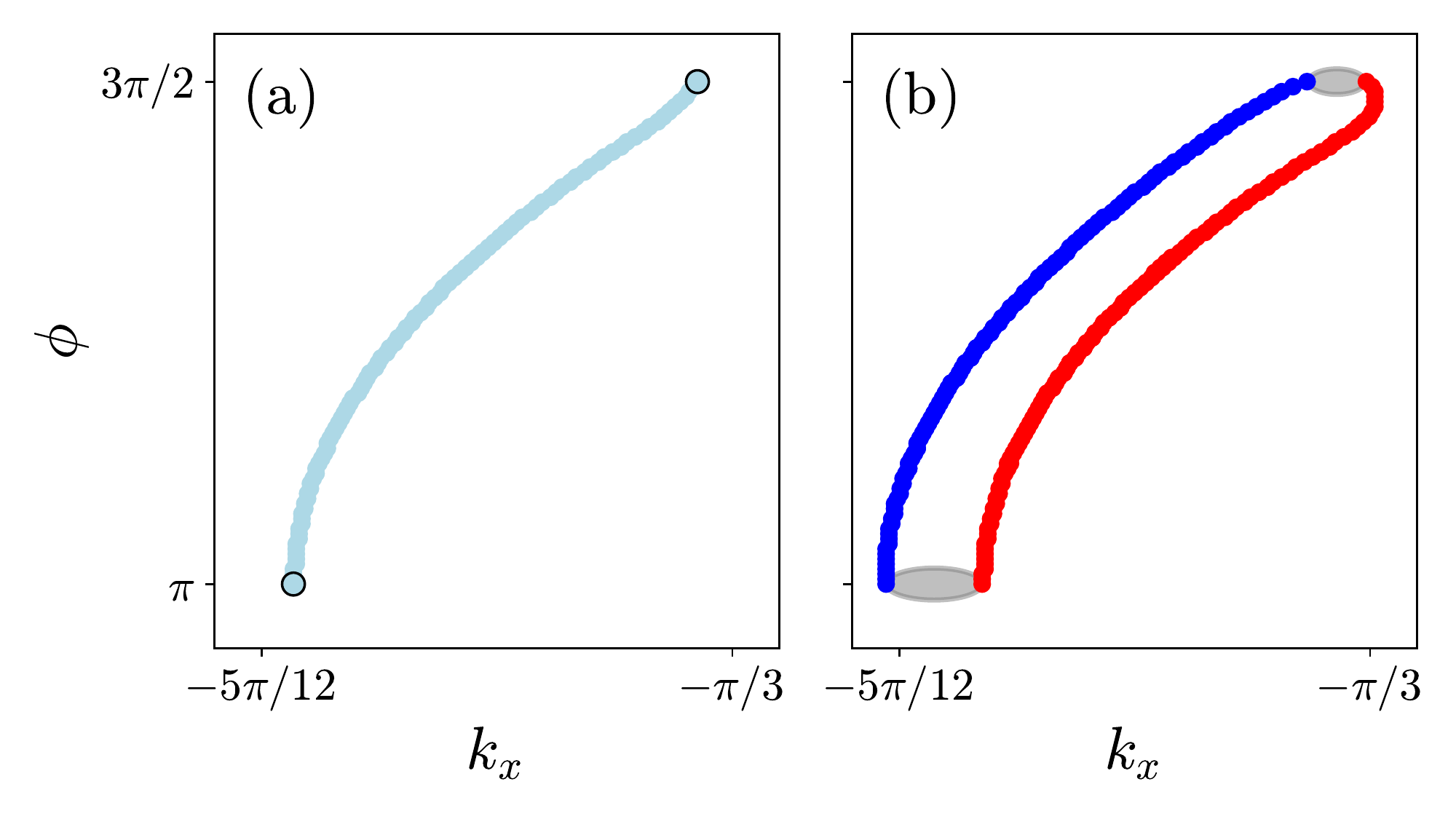}
\caption{(a) Fermi arcs traced on the $(k_x,\phi)$ plane by the
  gapless hinge modes of a rod extended along $x$ and with a $\z2$
  vertical termination, for the Fermi level at $E_{\rm F}=0$. (b)
  Same, but for $E_{\rm F}=0.2.$ The two elliptical discs indicate
  approximately the regions where the surface conduction bands move
  below $E=0.2$.  }
\figlab{fig8}
\end{figure}

\Fref{fig8} shows, for a $y_0$- and $\z2$-terminated rod, the locus of
points on the $(k_x,\phi)$ plane where the energy bands cross the
Fermi level.  In panel~(a), the Fermi level is at the
charge-neutrality point $E_{\rm F}=0$. In that case the locus of
points at $E_{\rm F}$ reduces to a four-fold degenerate Fermi arc in
$(k_x,\phi)$ space (the same on all hinges). In panel~(b) the Fermi
level has been shifted to $E_{\rm F}=0.2$, and as a result the Fermi
arc has split into a pair of two-fold degenerate arcs, where the
Kramers degeneracy again results from the fact that $\Lambda$ of
\eq{trs}, defined with respect to an inversion center in the middle of
the rod, commutes with the rod Hamiltonian.  The two Fermi arcs attach
tangentially to opposite sides of the two projected surface Fermi
surfaces, which have expanded from isolated points in panel~(a) to
finite disks [compare with \fref{fig2}(a)]. The way the Fermi arcs
close on adjacent hinges is analogous to the way they close on
opposite surfaces of a Weyl semimetal slab~\cite{armitage-rmp18}.

\subsection{Surface-hinge correspondence from slab Wannier bands}

We have seen how the quantized difference in AHC between two surfaces
dictates the occurrence of chiral modes on the connecting hinge.  In
this section we revisit this ``surface-hinge correspondence'' from the
viewpoint of the Wannier band structure of a slab.

\subsubsection{Hybrid Wannier representation}

Let us begin by reviewing the hybrid Wannier (HW) representation for a
$d$-dimensional insulating crystal~\cite{gresch-prb17}.  The idea is
to describe the valence states using functions that are
maximally-localized (Wannier-like) along one chosen crystallographic
direction $z$, and extended (Bloch-like) along the remaining $d\!-\!1$
directions. These HW functions $w_{ln}^\kk(\rr)$ are labeled by a
wavevector $\kk$ in the projected $(d\!-\!1)$-dimensional BZ, and by
two discrete indices $l$ and $n$; $l$ labels cells along $z$, and
$n=1,\ldots,J$ is an intracell index with $J$ being the number of
valence bands.  The HW centers
$z_{ln}(\kk)=\me{w_{ln}^\kk}{z}{w_{ln}^\kk}$ are organized into
``Wannier bands'' that are periodic in~$z$, with~$J$ bands per lattice
constant~$c$,
\beq
z_{ln}(\kk)=z_{0n}(\kk)+lc\,.
\eqlab{z-per}
\eeq
From now on, the HW centers will be written in units of the lattice
constant along the wannierization direction. Accordingly, we set $c=1$
in \eq{z-per}.

The Wannier band structure provides a very general means of
implementing the bulk-boundary
correspondence~\cite{fidkowski-prl11,neupert-bookchapter18}. Consider
for example a crystal in $d\!=\!3$ dimensions. When the boundary of
interest is a $z$-terminated surface, one inspects the bulk Wannier
bands $z_{ln}(k_x,k_y)$.  Under appropriate conditions to be specified
shortly, these can be smoothly deformed onto the surface energy bands
$E_n(k_x,k_y)$, so that the topological features of the two spectra
are in correspondence: any protected gapless modes in the surface
bands are reflected in the connectedness (or ``flow'') of the Wannier
bands~\cite{fidkowski-prl11,neupert-bookchapter18}.

For insulators with multiple occupied bands, the ability to make such
a smooth deformation depends on the choice of Wannier bands making up
a ``Wannier unit cell''~\cite{olsen-prb17}.  This is equivalent to the
choice of a Wannier gap separating one Wannier cell from the next
along $z$, or in the language of Ref.~\cite{khalaf-arxiv19}, ``fixing
the Wannier chemical potential.''  Specifically, if the Wannier unit
cell is repeated a large integer number of times along $z$, the
surface AHC at the top surface of the slab constructed in this way
must match that of the insulating surface in question, since if it
differs by an integer multiple of the quantum, a topological
obstruction prevents the smooth deformation.  In our case, the
correspondence is obvious: setting the Wannier gap at $z=0$ or $z=1/2$
is appropriate for the $z_0$- or $\z2$-terminated surface
respectively.  In general, however, a separate calculation may be
required to determine the correct choice of Wannier gap for a generic
insulating surface.

\subsubsection{Flow of surface-localized Wannier bands}

The surface-hinge correspondence can now be developed using closely
related methods.  To look for protected gapless modes on hinges
connecting $y$- and $z$-oriented surfaces, we examine the Wannier
bands $z_{ln}(k_x)$ of a $y$-terminated slab.  The interesting bands
are those whose HW functions reside near the surfaces.  If the flow of
these Wannier bands is such as to cross the Wannier gap appropriate to
the $z$-terminated surface of interest, then the $x$-directed $\yz$
hinges will host topologically protected gapless modes.  Identical
conclusions are reached by examining the bands $y_{ln}(k_x)$ of
$z$-terminated slabs.

\begin{figure}
\centering
\includegraphics[width=\columnwidth]{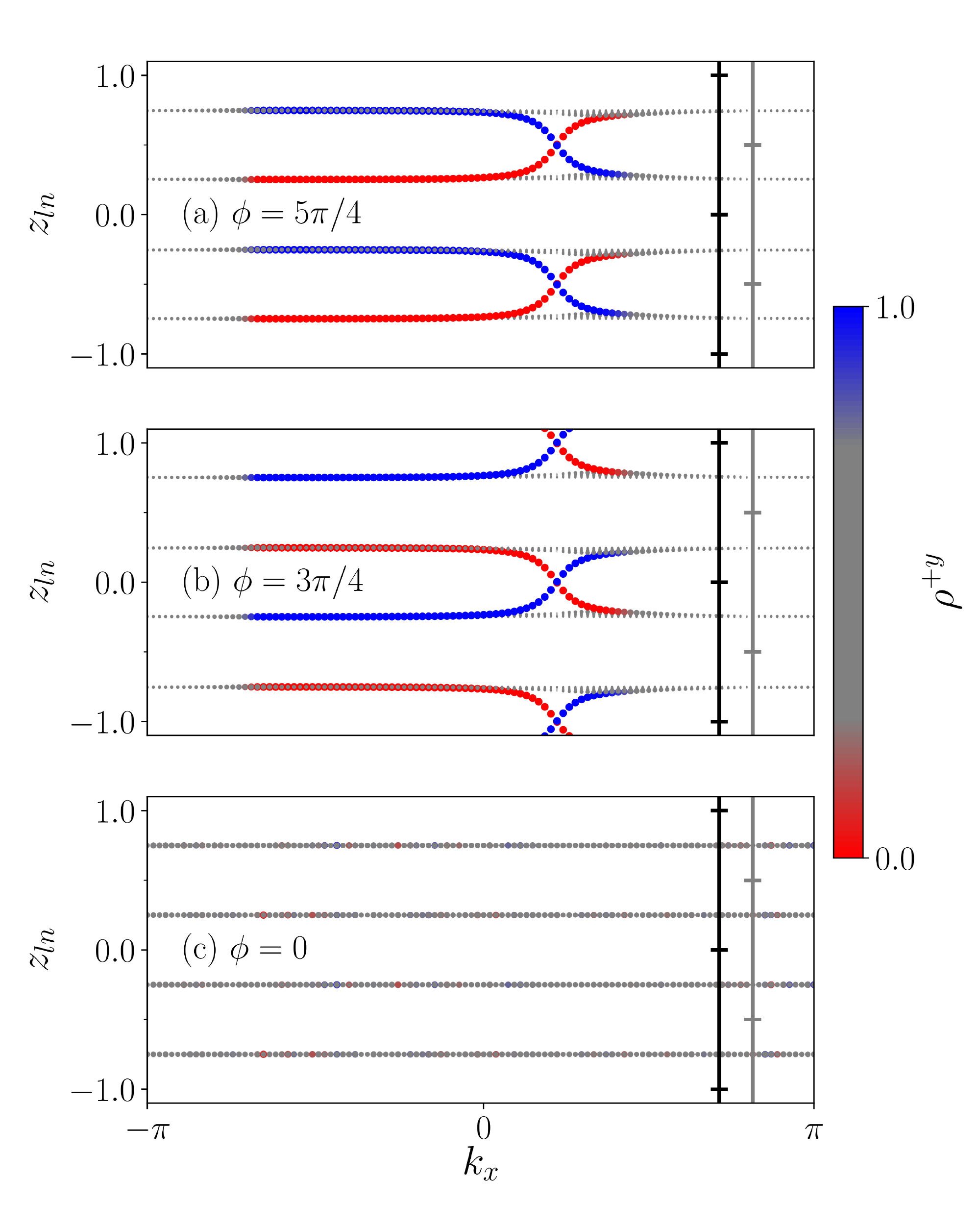}
\caption{Wannier bands $z_{ln}(k_x)$ of $y$-terminated slabs of the
  alternating Haldane model, at different $\phi$ values. The bands are
  color-coded according to the degree of localization on the $+\yhat$
  surface [\eq{rho}]: gray dots are modes extending along $y$ across
  the entire slab, and blue (red) dots are modes localized on the
  $+\yhat$ ($-\yhat$) surface; the degree of surface localization is
  also indicated by the size of the dots. Two types of cells are
  displayed in each panel: the ``$z_0$ cell'' with boundaries at
  $z=0\text{ mod 1}$ (in black), and the ``$\z2$ cell'' with
  boundaries at $z=1/2\text{ mod 1}$ (in gray).}
\figlab{fig9}
\end{figure}

\Fref{fig9} shows the bands $z_{ln}(k_x)$ of $y_0$-terminated slabs
with a thickness of 20 unit cells, calculated at $\phi=5\pi/4$,
$3\pi/4$, and $0$.  They are color-coded by the weight
\beq
\rho_n^{+y}(k_x)=\int_{+y}\left| w^{k_x}_{ln}(\rr)\right|^2\,d^3r
\eqlab{rho}
\eeq
of the HW functions in the half of the slab containing the $+\yhat$
surface, and for added clarity the degree of localization at the
surfaces is also indicated by the size of the dots.

Let us first examine the bands at $\phi=5\pi/4$ in panel~(a).  At
$k_x=-\pi$ they are evenly split into two narrow bulk-like groups, one
centered at $z=1/4\text{ mod 1}$ and another at $z=3/4\text{ mod
  1}$. Between them there is a ``$z_0$ gap'' centered at
$z=0\text{ mod 1}$, and a ``$\z2$ gap'' centered at
$z=1/2\text{ mod 1}$. As $k_x$ increases the two groups broaden
slightly, and one band detaches from each. The two detached bands
cross the $\z2$ gap in opposite directions, and as $k_x$ approaches
$\pi$ each merges with the bulk-like group from which the other came.
While crossing the gap, these two chiral bands become strongly
localized on opposite surfaces; this surface-localized flow across the
$\z2$ gap is maintained over the interval $\pi<\phi<3\pi/2$, signaling
the presence of protected gapless modes on $\yoz2$ hinges.\footnote{In
  \fref{fig9}(a), the Wannier band localized on the $+\yhat$ surface
  flows downward, in agreement with the negative chirality of the mode
  localized at the hinge between the $+\yhat$ and $+\zhat$ surfaces in
  \fref{fig7}.}
Conversely, the lack of flow on the $z_0$ gap indicates the absence of
such modes on $\yozo$ hinges over that interval.

The same logic applies to the other panels of \fref{fig9}.  In
panel~(b) the Wannier flow at $\phi=3\pi/4$ shifts to the $z_0$ gap
(and switches chirality), consistent with the fact that the $z_0$
termination is the one producing hinge modes (of the opposite
chirality), for $\pi/2<\phi<\pi$.  In panel~(c) both types of gaps are
devoid of chiral surface modes at $\phi=0$, reflecting the absence of
chiral hinge modes for $-\pi/2<\phi<\pi/2$ with either type of $z$
termination.

\begin{figure}
\centering
\includegraphics[width=\columnwidth]{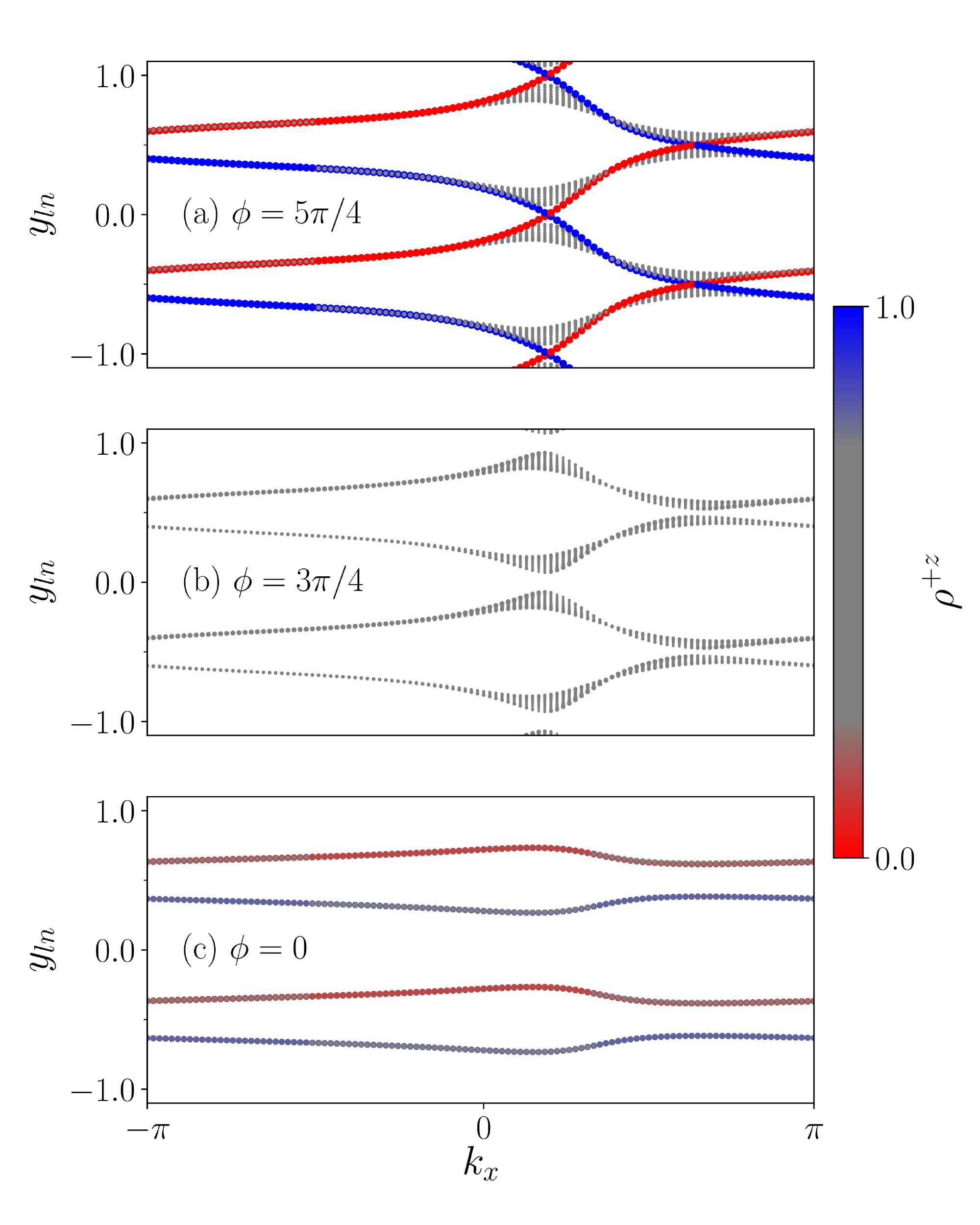}
\caption{Same as \fref{fig9}, but for the Wannier bands $y_{ln}(k_x)$
  of $\z2$-terminated slabs.}
\figlab{fig10}
\end{figure}

With the above procedure, we have been able to predict the existence
of gapless modes on both $\yozo$ and $\yoz2$ hinges from a single slab
calculation (at each $\phi$). This is somewhat unexpected, given that
the surface-AHC approach of \sref{surf-AHC} required three separate
slab calculations to gather the same information. It should be noted,
however, that the HW-based procedure only works when the choice of
Wannier gap corresponding to the $z$-terminated surface of interest is
known, whereas the surface-AHC approach can be applied directly to
arbitrary insulating surfaces.

The protected modes on $\yz$ hinges can also be deduced from the
Wannier spectrum $y_n(k_x)$ of $z$-terminated slabs, but this requires
two slab calculations instead of one (one for each type of $z$
termination).  This is illustrated in \fref{fig10} for the case of
$\yoz2$ hinges, using $\z2$-terminated slabs.  As expected, Wannier
flow is present on the $y_0$ gap at $\phi=5\pi/4$ but not at
$\phi=3\pi/4$ or at $\phi=0$.

\subsubsection{Interpretation in terms of charge pumping at the
  surface}

\begin{figure}
\centering
\includegraphics[width=\columnwidth]{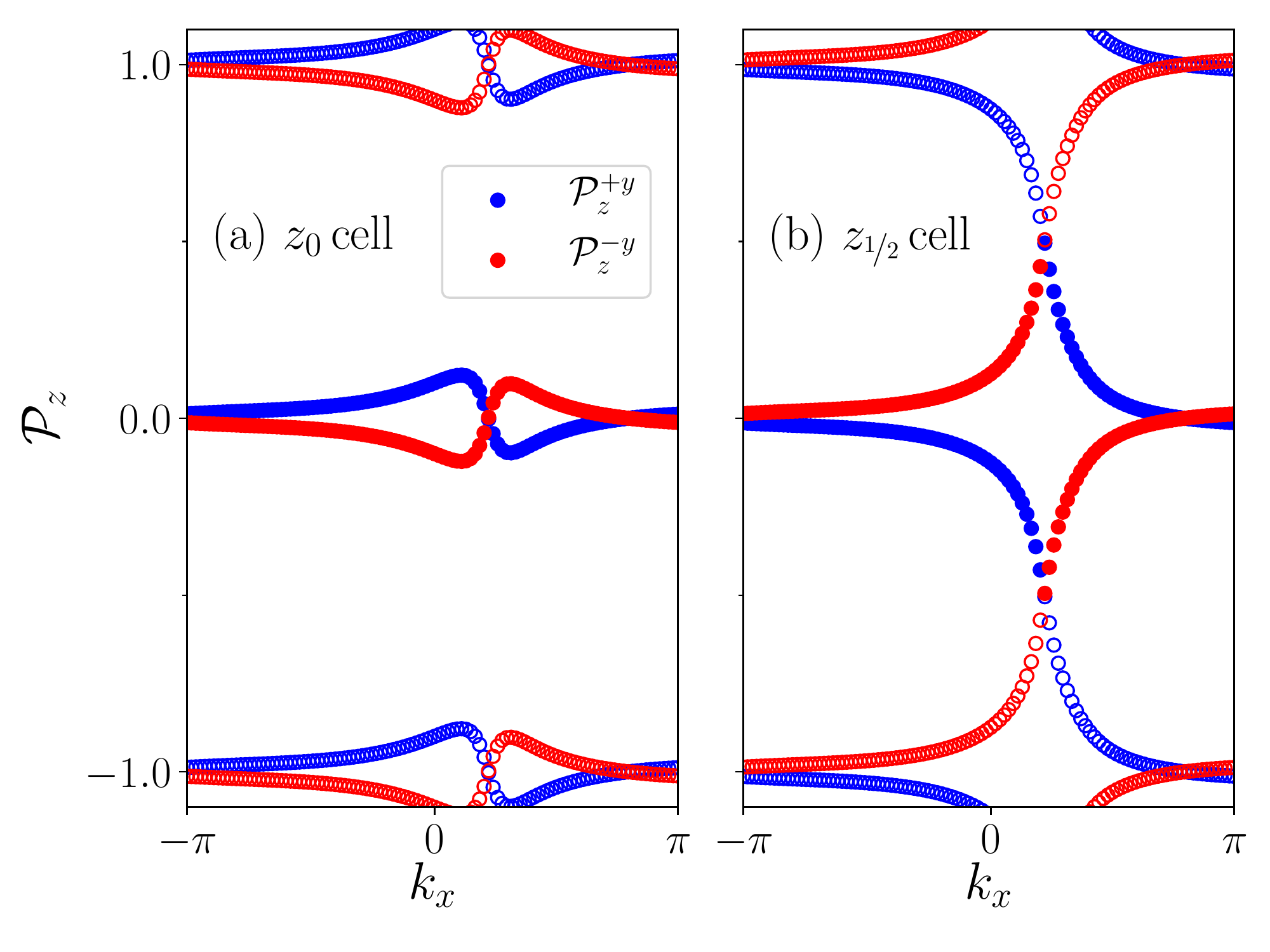}
\caption{Pumped charge ${\cal P}_z(k_x)$ of \eq{surface-p}, in units
  of $e$, for a $y$-terminated slab at $\phi=5\pi/4$.  Blue (red)
  curves denote results for the $+\yhat$ ($-\yhat$) surface.  In (a)
  and (b), \eq{surface-p} is evaluated using the $z_0$ and $\z2$ cells
  shown in \fref{fig9}, respectively. Filled dots correspond to the
  home cell ($[0,1]$ in (a) and $[-1/2,1/2]$ in (b)), and open dots
  correspond to the cells immediately above and below.}
\figlab{fig11}
\end{figure}

The information in \ffrefs{fig9}{fig10} about the topology of $\yz$
hinges can be presented concisely in the language of charge
pumping. Let us describe the procedure for the case of \fref{fig9},
where the slab is terminated along~$y$. Following
Ref.~\cite{benalcazar-prb17}, we assign to the $+\yhat$ surface a
quantity with units of charge defined as
\beq
{\cal P}_z^{+y}(k_x)=-\frac{e}{c}\sum_n\,z_{0n}(k_x)
\rho^{+y}_n(k_x)\,,
\eqlab{surface-p}
\eeq
where $\rho^{+y}_n$ is given by \eq{rho}.  For a given value of $k_x$,
this quantity is a measure of the charge pumped along~$z$ on the
$+\yhat$ edge of the ribbon (finite in $y$, infinite in $z$) described
by $H_{\rm slab}(k_x)$.  However, its physical interpretation is
rather subtle.  For example, consider a weak electric field along $x$
that acts for one Bloch period. In this case, the continuous change in
${\cal P}_z^{+y}$ as $k_x$ increases by $2\pi$ is quantized in units
of $e$, and it describes the $+\zhat$-directed flow of current on the
$+\yhat$ surface {\it relative} to the $-\yhat$-directed current on
the $+\zhat$ surface (that is, the quantized difference $\Delta n$ in
surface AHC). Note that \eq{surface-p} depends on the choice of
Wannier unit cell, and again the answer will only be correct if that
cell is chosen correctly for the $+\zhat$-terminated surface of
interest.

In \fref{fig11}, ${\cal P}_z^{+y}$ is plotted at $\phi=5\pi/4$ for two
different cell choices.  In panel~(a), the black $z_0$ cell in
\fref{fig9} was used.  Since in \fref{fig9}(a) the chiral Wannier band
localized on the $+\yhat$ surface does not cross the boundaries of
that cell, ${\cal P}_z^{+y}$ does not exhibit flow as a function of
$k_x$, indicating that no protected gapless modes are present on the
$\yozo$ hinges.  In \fref{fig11}(b) the calculation was repeated using
the gray $\z2$ cell in \fref{fig9}. Now the surface-localized band
does cross the cell boundaries, and as a result ${\cal P}_z^{+y}$
exhibits flow as a function of $k_x$ (when viewed as a continuous but
multivalued function), indicating the presence of gapless modes on the
$\yoz2$ hinges. Also shown in \fref{fig11} is ${\cal P}_z^{-y}$,
obtained by replacing $\rho_n^{+y}$ with $\rho_n^{-y}=1-\rho_n^{+y}$
in \eq{surface-p}.

\section{Summary and outlook}
\seclab{summary}

We have shown that gapless modes appear naturally on the hinges of 3D
insulators undergoing an axion pumping cycle.  The basic idea is
illustrated in \ffrefs{fig1}{fig2}. When a surface is introduced in
the system, the valence and conduction surface bands must exhibit at
least one nodal touching along the cycle. If, as is generically the
case, those band touchings occur on adjacent surfaces at different
values of the pumping parameter $\phi$, then the connecting hinge will
host chiral modes over the intervening $\phi$ range. Those modes are
boundary manifestations of the second Chern number characterizing the
axion pump, and they can be viewed as Fermi arcs in the BZ of the 2D
hinge connecting the 3D surfaces of a 4D sample with
$(k_x,k_y,k_z,\phi)$ reciprocal space.

Note that at any given value of $\phi$, the appearance of 1D modes on
the hinges of the 3D crystal represents an ``extrinsic'' higher-order
topological phase in the language of Refs.~\cite{geier-prb18}] and
\cite{trifunovic-prx19}, since the bulk is topologically trivial and
hinge modes are not required.  Instead, the presence of Fermi arc
states is generically required on the 2D surfaces of the 4D
second-Chern insulator, thus representing ``intrinsic'' topology when
the system is viewed from the standpoint of the global
$(k_x,k_y,k_z,\phi)$ parameter space.

We have exemplified these behaviors by means of a tight-binding model,
but the same methodology could easily be applied in the framework of
\textit{ab initio} calculations. However, it remains a major challenge
to devise a physical mechanism leading to the adiabatic pumping of
axion coupling in a real material.

Alternatively, it may be possible to demonstrate axion pumping
behavior in other settings such as photonic crystals, ultracold atoms,
or electrical circuits. The physics of second-Chern insulator is
already being explored in such
systems~\cite{ozawa-pra16,zilberberg-nature18,lu-natcomms18,zhang-prb19,price-prl15,price-prb16,lohse-nature18,ezawa-prb19},
and we hope that the present work may inspire future efforts towards
the observation of the associated topological hinge states.

\acknowledgments{Work by T.O.\ was funded by the Danish Independent
  Research Foundation, Grant number 6108-00464B.  Work by T.R.\ was
  supported by the Forschungsstipendium Grant No.\ RA 3025/1-1 from
  the Deutsche Forschungsgemeinschaft.  Work by D.V.\ was supported by
  National Science Foundation Grant DMR-1954856. Work by I.S.\ was
  supported by Grant No.\ FIS2016-77188-P from the Ministerio de
  Ciencia e Innovación (Spain).}


%

\end{document}